\documentclass[letterpaper,twocolumn,prl,aps,superscriptaddress,amsmath,amssymb,floatfix]{revtex4-1}
\usepackage{newtxtext}
\usepackage{newtxmath}
\usepackage[latin9]{inputenc}
\usepackage{color}
\usepackage{amsmath}
\usepackage{amssymb}
\usepackage{graphicx}
\graphicspath{{figures/}}
\usepackage{esint}
\usepackage{siunitx}
\usepackage[unicode=true,
 bookmarks=true,bookmarksnumbered=false,bookmarksopen=false,
 breaklinks=false,pdfborder={0 0 1},backref=false,colorlinks=true]
 {hyperref}
\hypersetup{
 linkcolor=magenta,urlcolor=blue,citecolor=blue,pdfstartview={FitH},hyperfootnotes=false}
\makeatletter

\usepackage{textcomp}
\usepackage{epstopdf}

\pdfpageheight\paperheight
\pdfpagewidth\paperwidth

\@ifundefined{textcolor}{}{%
 \definecolor{BLACK}{gray}{0}
 \definecolor{WHITE}{gray}{1}
 \definecolor{RED}{rgb}{1,0,0}
 \definecolor{GREEN}{rgb}{0,1,0}
 \definecolor{BLUE}{rgb}{0,0,1}
 \definecolor{CYAN}{cmyk}{1,0,0,0}
 \definecolor{MAGENTA}{cmyk}{0,1,0,0}
 \definecolor{YELLOW}{cmyk}{0,0,1,0}
}

\usepackage{xcolor}\usepackage{soul}
\newcommand{\ket}[1]{\ensuremath{\left|#1\right\rangle}}

\definecolor{blue}{rgb}{0,0,1}
\definecolor{red}{rgb}{1,0,0}
\definecolor{green}{rgb}{0,1,0}

\usepackage{soul}

\makeatother

\begin{document}

\title{A scalable chip-interfaced single-photon source array based on 50 individually addressable neutral atoms}

\author{Ya-Dong~Hu}
\thanks{These authors contributed equally to this work.}
\affiliation{Laboratory of Quantum Information, University of Science and Technology of China, Hefei 230026, China.}
\affiliation{CAS Center for Excellence in Quantum Information and Quantum Physics,
University of Science and Technology of China, Hefei 230026,
China.}

\author{Tian-Yang~Zhang}
\thanks{These authors contributed equally to this work.}
\affiliation{Laboratory of Quantum Information, University of Science and Technology of China, Hefei 230026, China.}
\affiliation{CAS Center for Excellence in Quantum Information and Quantum Physics,
University of Science and Technology of China, Hefei 230026,
China.}

\author{Dong-Qi~Ma}
\thanks{These authors contributed equally to this work.}
\affiliation{Laboratory of Quantum Information, University of Science and Technology of China, Hefei 230026, China.}
\affiliation{CAS Center for Excellence in Quantum Information and Quantum Physics,
University of Science and Technology of China, Hefei 230026,
China.}

\author{Yi-Chen~Zhang}
\affiliation{Laboratory of Quantum Information, University of Science and Technology of China, Hefei 230026, China.}
\affiliation{CAS Center for Excellence in Quantum Information and Quantum Physics,
University of Science and Technology of China, Hefei 230026,
China.}

\author{Liang~Chen}
\affiliation{Laboratory of Quantum Information, University of Science and Technology of China, Hefei 230026, China.}
\affiliation{CAS Center for Excellence in Quantum Information and Quantum Physics,
University of Science and Technology of China, Hefei 230026,
China.}

\author{Wen-Yi~Zhu}
\affiliation{Laboratory of Quantum Information, University of Science and Technology of China, Hefei 230026, China.}
\affiliation{CAS Center for Excellence in Quantum Information and Quantum Physics,
University of Science and Technology of China, Hefei 230026,
China.}

\author{Hong-Jie~Fan}
\affiliation{Laboratory of Quantum Information, University of Science and Technology of China, Hefei 230026, China.}
\affiliation{CAS Center for Excellence in Quantum Information and Quantum Physics,
University of Science and Technology of China, Hefei 230026,
China.}

\author{Yan-Lei Zhang}
\affiliation{Laboratory of Quantum Information, University of Science and Technology of China, Hefei 230026, China.}
\affiliation{CAS Center for Excellence in Quantum Information and Quantum Physics,
University of Science and Technology of China, Hefei 230026,
China.}

\author{Zhu-Bo~Wang}
\email{zbwang@ustc.edu.cn}
\affiliation{Laboratory of Quantum Information, University of Science and Technology of China, Hefei 230026, China.}
\affiliation{CAS Center for Excellence in Quantum Information and Quantum Physics,
University of Science and Technology of China, Hefei 230026,
China.}

\author{Gang~Li}
\email{gangli@sxu.edu.cn}
\affiliation{Key Laboratory of Quantum Optics and Quantum Optics Devices, Institute of Opto-Electronics, Shanxi University, Taiyuan 030006, China}
\affiliation{Collaborative Innovation Centre of Extreme Optics, Shanxi University, Taiyuan 030006, China}

\author{Xi-Feng~Ren}
\email{renxf@ustc.edu.cn}
\affiliation{Laboratory of Quantum Information, University of Science and Technology of China, Hefei 230026, China.}
\affiliation{CAS Center for Excellence in Quantum Information and Quantum Physics,
University of Science and Technology of China, Hefei 230026,
China.}
\affiliation{Hefei National Laboratory, Hefei 230088, China.}

\author{Guang-Can~Guo}
\affiliation{Laboratory of Quantum Information, University of Science and Technology of China, Hefei 230026, China.}
\affiliation{CAS Center for Excellence in Quantum Information and Quantum Physics,
University of Science and Technology of China, Hefei 230026,
China.}
\affiliation{Hefei National Laboratory, Hefei 230088, China.}

\author{Chang-Ling~Zou}
\email{clzou321@ustc.edu.cn}
\affiliation{Laboratory of Quantum Information, University of Science and Technology of China, Hefei 230026, China.}
\affiliation{CAS Center for Excellence in Quantum Information and Quantum Physics,
University of Science and Technology of China, Hefei 230026,
China.}
\affiliation{Hefei National Laboratory, Hefei 230088, China.}

\date{\today}

\begin{abstract}
{\bfseries\boldmath Scalable arrays of identical single-photon sources are a central resource for photonic quantum information processing, quantum networks and quantum metrology. Neutral atoms provide intrinsically identical emitters that can be assembled and rearranged in optical tweezers, but a many-channel fiber interface to individually trapped atoms has remained a major technical challenge. Here we demonstrate a chip-interfaced single-photon source array based on 50 individually addressable $^{87}\mathrm{Rb}$ atoms. A glass waveguide fan-out converts the \SI{5}{\micro m} pitch of the optical-tweezer array to the \SI{127}{\micro m} pitch of a commercial fiber array, mapping each atom to its own waveguide, fiber and single-photon detector. We resolve all 50 channels with an average nearest-neighbor cross-talk of $0.4\%$ and a uniform insertion loss of \SI{2.9}{dB}, and verify single-photon emission with $g^{(2)}(0)=0.29$, presently limited by detector dark counts and residual cooling-light scattering. Combining per-channel atom discrimination, rearrangement and reservoir replenishment, we prepare source subarrays of up to 24 atoms with a $93\%$ fill fraction. For small target numbers, atom loss is repaired from the reservoir at the detection-limited rate of \SI{118}{Hz}. We further fabricate a 784-channel waveguide chip, showing that the photonic interface can be extended well beyond the present number. This architecture establishes a fiber-native neutral-atom platform for larger arrays of identical single-photon sources.}
\end{abstract}
\maketitle

\noindent \textbf{\large{}Introduction}{\large\par}

\noindent Arrays of identical single-photon sources are an important building block for photonic quantum computing and simulation~\cite{OBrien2009,Takeda2019,quantumsimulator,Lindner2009}, quantum networks~\cite{Kimble2008} and quantum-enhanced metrology~\cite{Giovannetti2011}. Beyond the brightness and purity of an individual emitter, a practical source array requires mutual emitter identity, deterministic spatial addressing, stable routing into fiber or chip-based optical circuits, and continuous operation with low dead time~\cite{photonsourceReview,Lodahl2017,MeyerScott2020}.

Solid-state emitters and nonlinear photon-pair sources have enabled substantial progress toward scalable photonic sources~\cite{Kwiat1995,photonsourceReview,Senellart2017}. However, when many independent sources are required, these platforms typically rely on spectral tuning, active multiplexing or post-selection, because each emitter or nonlinear process is affected by its local environment~\cite{Aharonovich2016,MeyerScott2020}. Neutral atoms offer a complementary route because atoms of the same isotope have transition frequencies fixed by atomic structure, as demonstrated by the interference of photons emitted from independently trapped atoms~\cite{Beugnon2006}, while optical-tweezer techniques now allow atoms to be trapped, rearranged and individually addressed in large arrays~\cite{Schlosser2001,Barredo2016,Endres2016,Browaeys2020,Schymik2020,Pause2024}. These properties make neutral atoms a natural platform for reproducible single-photon source arrays~\cite{Darquie2005}.

The missing element is a scalable photonic interface. Most neutral-atom-array experiments collect fluorescence in free space and image it onto an EMCCD or sCMOS camera~\cite{Endres2016,Barredo2016,Bluvstein2024}. Although this approach is efficient for detecting atom occupation, it does not provide independent, fiber-routed optical channels from individual atoms. Direct coupling to fibers, waveguides or cavities can provide individually resolved channels for atom control or photon routing~\cite{Lodahl2017,Tiecke2014,Dordevic2021,Menon2024,Li2025FiberArray,Ma2026Volcano}, but fiber-coupled atomic single-photon sources have so far been restricted to one or a few emitters~\cite{Darquie2005,McKeever2004}. A central difficulty is the pitch mismatch, as the several-micrometer spacing of an optical-tweezer array cannot be matched directly to the hundred-micrometer spacing of standard fiber and detector arrays without compromising single-mode coupling.

Here we address this problem with a glass waveguide fan-out that converts the microscopic pitch of an optical-tweezer array to the standard pitch of a commercial fiber array. The interface maps each atom one-to-one onto a waveguide channel, a fiber and a single-photon detector. We realize a source array of 50 individually addressable $^{87}\mathrm{Rb}$ atoms, quantify the channel mapping, insertion loss and cross-talk, and verify the single-photon character of the collected light. By combining per-channel atom discrimination with rearrangement and reservoir replenishment, we further prepare source subarrays of up to 24 atoms and restore them after atom loss at a detection-limited reset rate of up to \SI{118}{Hz}. Finally, we fabricate a 784-channel chip based on the same architecture. These results provide, to our knowledge, the largest fiber-interfaced neutral-atom single-photon source array and establish a scalable, fiber-native interface for neutral-atom photon sources.

\begin{figure*}
\begin{centering}
\includegraphics[width=1\textwidth]{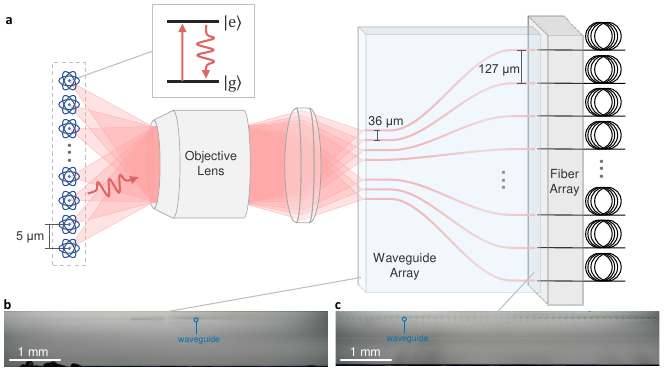}
\par\end{centering}
\caption{\textbf{A chip-interfaced waveguide fan-out for neutral-atom single-photon source arrays.} \textbf{a}, Optical layout. A one-dimensional array of $^{87}\mathrm{Rb}$ atoms (left) is held in optical tweezers; each atom acts as a two-level emitter ($\ket{e}\!\rightarrow\!\ket{g}$). Their fluorescence is collected by an objective and relayed onto the input facet of the waveguide array, where the magnified atom spacing matches the \SI{36}{\micro\meter} waveguide pitch. The chip then fans the channels out to a \SI{127}{\micro\meter} pitch at the output facet, matching a commercial fiber array. \textbf{b}\&\textbf{c}, Micrographs of the input (\textbf{b}) and output (\textbf{c}) facets of the chip, recorded at the same magnification. The blue circle marks a single waveguide. Waveguides across the row differ in apparent brightness and contrast because the facet is not uniformly illuminated during imaging.}
\label{Fig1}
\end{figure*}

\begin{figure*}[!t]
\begin{centering}
\includegraphics[width=1\textwidth]{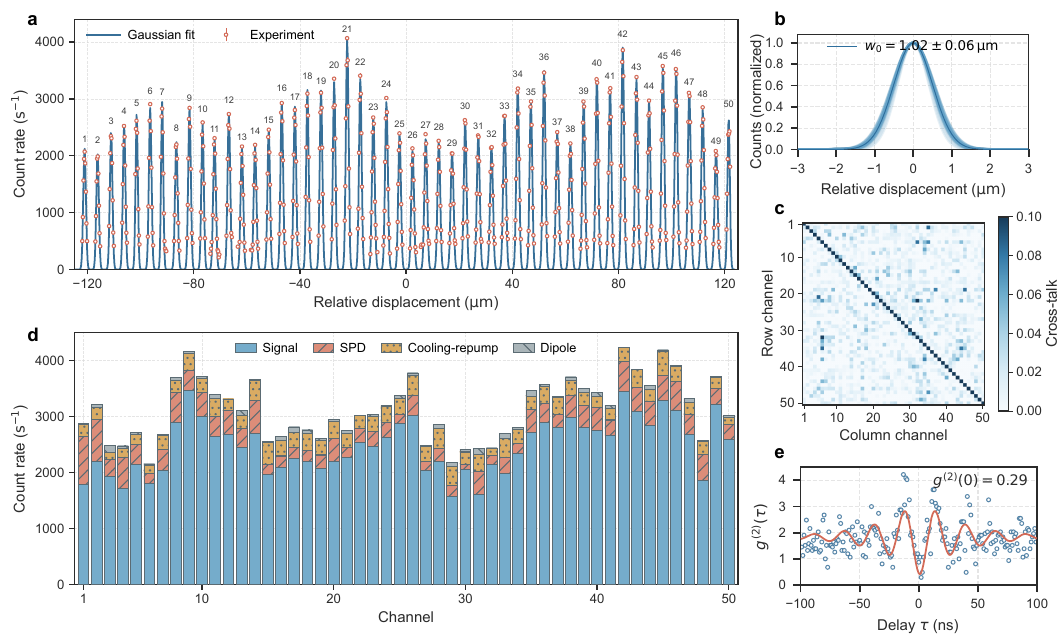}
\par\end{centering}
\caption{\textbf{One-to-one channel mapping and simultaneous operation of 50 neutral-atom single-photon sources.} \textbf{a}, Fluorescence collected by each of the 50 channels as a single atom is translated across the focal plane, the displacement being controlled by the radio-frequency (RF) drive of the tweezer acousto-optic deflector (AOD). The scan resolves 50 peaks, each corresponding to one waveguide channel (labeled $1$-$50$). \textbf{b}, Collection mode line shape. After removing their relative offsets and normalizing, the individual peaks (light blue) overlap; the averaged response (dark blue) is Gaussian with a waist $w_{0}=\SI{1.02}{\micro m}$, set by the fluorescence collection mode at the atom plane. \textbf{c}, Measured $50\times50$ cross-talk matrix. Column $i$ is obtained with only tweezer $i$ illuminated: all 50 channels are read out and the shots are post-selected on trap $i$ being occupied. The element $c_{ji}=\langle n_j\rangle_i/\langle n_i\rangle_i$ is the mean count rate in channel $j$ normalized to that in the illuminated channel $i$. The average nearest-neighbor cross-talk is $0.4\%$. \textbf{d}, fluorescence for all 50 channels, decomposed into the atomic signal, single-photon-detector (SPD) dark counts, the cooling/repump background and the dipole-trap background. \textbf{e}, Second-order correlation $g^{(2)}(\tau)$ of the photons emitted into channel 26 with $g^{(2)}(0)=0.29$ (below the $0.5$ threshold), confirming single-photon emission. Circles are normalized coincidence counts in \SI{1}{\nano \second} bins, and the solid line is a fit to the two-level optical Bloch equations.}
\label{Fig2}
\end{figure*}

\begin{figure*}[!t]
\begin{centering}
\includegraphics[width=0.9\textwidth]{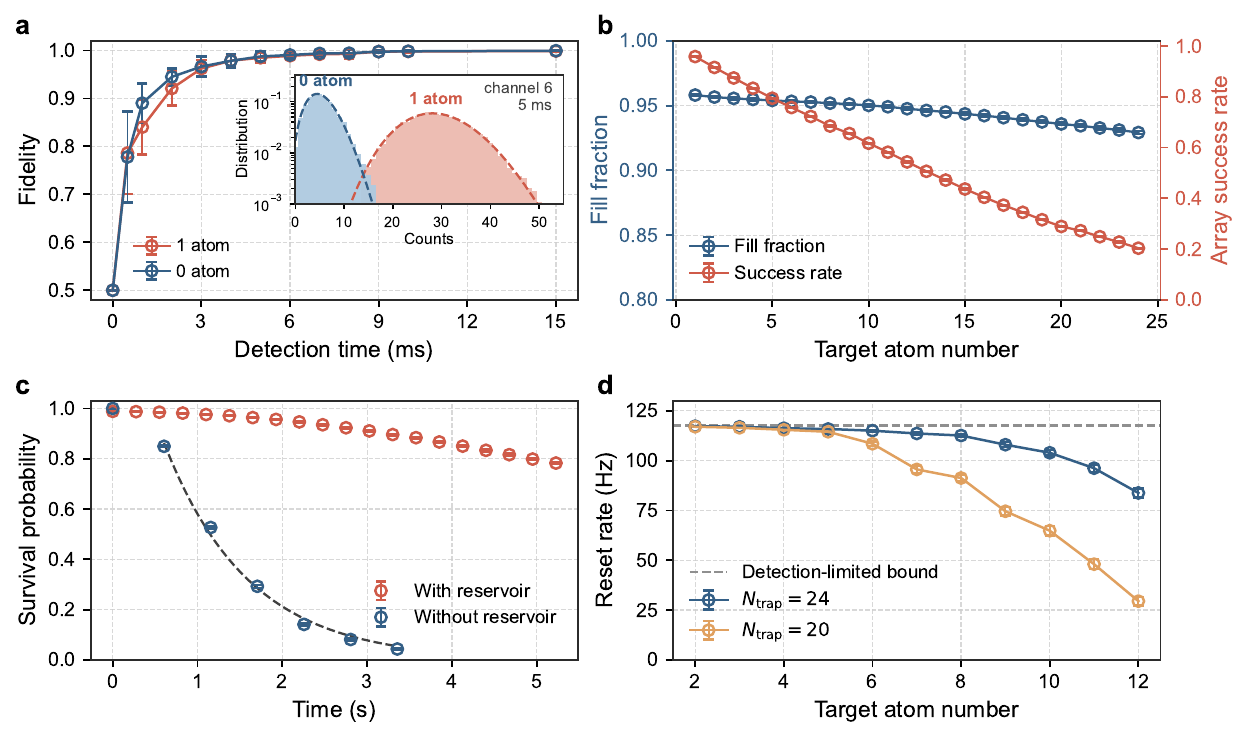}
\par\end{centering}
\caption{\textbf{State readout, deterministic loading, and stable operation of the source array.} \textbf{a}, Single-atom readout fidelity versus detection time, approaching near-unity at longer integration. The inset shows the normalized photon-count distribution for channel 6 over a \SI{5}{ms} window, resolving the 0-atom and 1-atom states with a readout fidelity of $98.6\%$. \textbf{b}, Average single-trap fill fraction after rearrangement (left axis, blue) and whole-array preparation success rate (right axis, red) as a function of the target atom number $N$, starting from an initial array of 48 traps. \textbf{c}, Single-atom survival probability versus holding time. The red data were obtained with 10 target atoms prepared by rearrangement in a 24-trap array. Atoms occupying traps outside the 10 target sites form a reservoir and replenish atoms lost from the target array throughout the holding sequence. The dashed line is an exponential fit to the without-reservoir data, excluding the initial plateau. Reservoir replenishment extends the effective $1/e$ lifetime from $\approx\SI{1}{s}$ to $\approx\SI{10}{\second}$. \textbf{d}, Array reset rate at which a fresh defect-free array of $N$ sources is restored by fast per-channel atom discrimination, rearrangement and reservoir replenishment--versus target atom number $N$ for trapping-array sizes $N_{\mathrm{trap}}=20$ and $24$. The gray dashed line marks the detection-limited upper bound of \SI{118}{Hz}.}
\label{Fig3}
\end{figure*}

\begin{figure*}[!t]
\begin{centering}
\includegraphics[width=0.78\textwidth]{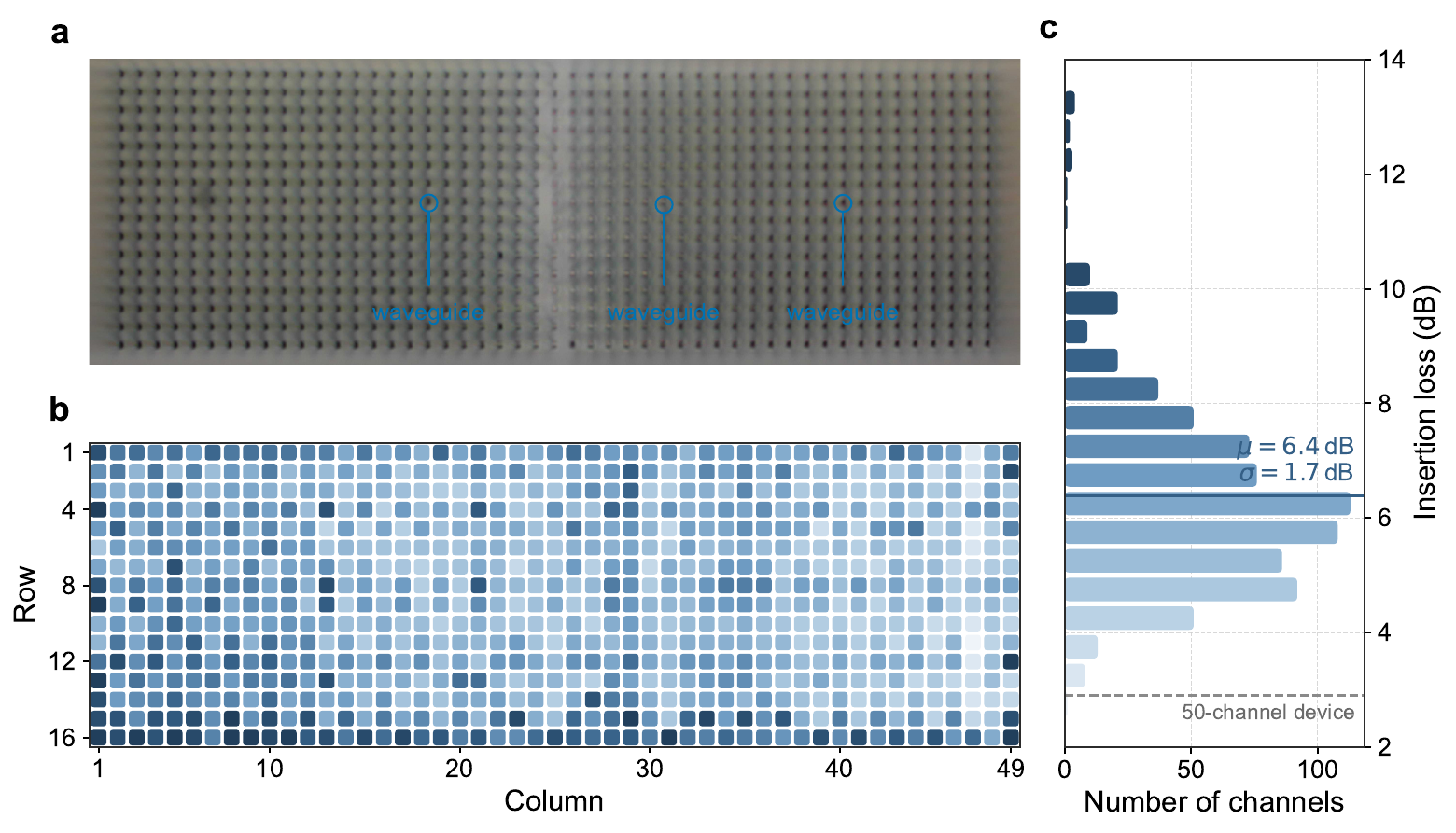}
\par\end{centering}
\caption{\textbf{Insertion-loss characterization of a 784-channel waveguide array.} \textbf{a}, Optical micrograph of the input facet, where the channels form a 16-row by 49-column array with a \SI{36}{\micro m} pitch. At the opposite facet, the channels are rearranged into a $4\times196$ array with a \SI{127}{\micro m} pitch, matching a two-dimensional fiber array. The blue circles mark three individual waveguides. Variations in the apparent brightness and contrast of the channels across the micrograph arise from non-uniform illumination during imaging. \textbf{b}, Spatial map of the insertion loss of all 784 channels (196 groups of four waveguides), arranged according to their physical positions in \textbf{a}. Each rounded square represents one channel and is colored by $-10\log_{10}T$, where $T$ is the measured fractional transmittance. \textbf{c}, Distribution of the per-channel insertion loss; the bars use the same color mapping as in \textbf{b}. The solid line marks the mean, $\mu=\SI{6.4}{dB}$, and the standard deviation is $\sigma=\SI{1.7}{dB}$. One channel with $T<1\%$ is retained in \textbf{b} but excluded from the histogram and statistics. The gray dashed line marks the mean insertion loss of the 50-channel device, \SI{2.9}{dB}.}
\label{Fig4}
\end{figure*}

\smallskip{}

\noindent \textbf{\large{}Results}{\large\par}

\noindent \textbf{A waveguide fan-out for neutral-atom source arrays}

\noindent The source architecture is shown in Fig.~\ref{Fig1}(a). Its central element is a glass waveguide fan-out~\cite{Ma2026Volcano,Davis1996,Marshall2009,Meany2015,Tan2021,Corrielli2021} that connects the microscopic length scale of a neutral-atom array to the standard pitch of a fiber array. A one-dimensional optical-tweezer array of $^{87}\mathrm{Rb}$ atoms is generated by driving a single-axis acousto-optic deflector (AOD) with a multi-tone RF signal at a trap wavelength of \SI{852}{nm}. The diffracted beams are focused by a home-built objective ($\mathrm{NA}=0.4$), forming traps with a \SI{5}{\micro m} pitch. The same objective collects the atomic fluorescence, which is relayed to the input facet of the waveguide chip. The imaging system first magnifies the atomic pitch from \SI{5}{\micro m} to \SI{36}{\micro m}, then the waveguide array performs the remaining conversion to a \SI{127}{\micro m} pitch at the output facet, realizing the optical channel mapping to match a commercial fiber array~\cite{Ma2026Volcano}. Figures~\ref{Fig1}(b) and \ref{Fig1}(c) show micrographs of the input and output facets, respectively. Each fiber is connected to an individual single-photon detector (SPD), thereby forming a one-to-one chain from atom to waveguide, fiber and electronic detection channel.

The fan-out decouples mode matching from pitch matching, two functions that a purely free-space interface must set with a single imaging magnification. Efficient single-photon collection requires a well-defined spatial mode, because mode purity is essential for high-visibility interference and coherent routing~\cite{Hong1987,Beugnon2006,Lodahl2017}. If the atom array were magnified directly to the \SI{127}{\micro m} fiber pitch, the fluorescence spot at each output would become much larger than the mode field of a single-mode fiber, reducing the coupling efficiency. In contrast, the guided-wave fan-out changes the channel pitch after the light has been coupled into waveguides, preserving the collected mode size while providing fiber-array-compatible outputs.

\vspace{6pt}
\noindent \textbf{One-to-one mapping of 50 atom-waveguide channels}

\noindent To calibrate the atom-waveguide mapping, we translated a single atom across the focal plane by scanning the RF frequency applied to the tweezer AOD, while recording the fluorescence collected by all waveguide channels. The RF-frequency-to-displacement conversion of the optical tweezer is calibrated by imaging. The resulting map resolves 50 separated peaks [Fig.~\ref{Fig2}(a)], each corresponding to one waveguide channel. This verifies that the 50 detection channels independently address 50 spatial positions of the atom array. After the relative offsets of different channels are removed, the normalized traces overlap well, and the averaged response is fitted by a Gaussian profile with a waist of \SI{1.02(6)}{\micro\meter} [Fig.~\ref{Fig2}(b)], where the uncertainty is the channel-to-channel standard deviation. This waist represents the collection mode at the atom plane and remains small compared with the \SI{5}{\micro m} tweezer spacing, ensuring that each channel predominantly collects fluorescence from one trap.

The passive photonic interface was further characterized by injecting light backwards from the fiber array and collecting the transmitted light before the input facet. This measurement includes fiber-array-to-waveguide coupling and waveguide propagation loss, giving an average insertion loss of \SI{2.9}{\decibel} with a channel-to-channel standard deviation of \SI{0.2}{\decibel}. The optical cross-talk was measured separately, with atoms: the tweezers were illuminated one at a time, all 50 channels were read out, and the shots were post-selected on the illuminated trap being occupied. The cross-talk matrix element is defined as $c_{ji}=\langle n_j\rangle_i/\langle n_i\rangle_i$, where $\langle\cdot\rangle_i$ denotes the mean over shots (background removed) with only tweezer $i$ illuminated and occupied, and the full $50\times50$ matrix is shown in Fig.~\ref{Fig2}(c). The average
nearest-neighbor cross-talk is $0.4\%$, with a maximum of $5.8\%$. Fluorescence from one atom is therefore routed predominantly to its corresponding channel, and the focused waveguide modes are sufficiently separated at the atom plane.

When all sources are operated together, the integrated fluorescence of the 50 channels can be decomposed into the atomic signal and the background contributions from SPD dark counts, cooling/repump light and dipole-trap light [Fig.~\ref{Fig2}(d)]. The residual background is mainly attributed to detector dark counts, indicating that lower-noise detectors can directly improve the readout contrast and reduce the detection time. To verify the photon statistics, we measured the second-order correlation function of channel 26 and obtained $g^{(2)}(0)=0.29$ [Fig.~\ref{Fig2}(e)], below the $0.5$ threshold for single-photon emission~\cite{Kimble1977}. This measurement used two separate cryogenic single-photon detectors with lower dark-count rate than those used in Fig.~\ref{Fig2}(d). The solid line is a fit to the two-level optical Bloch equations for resonance fluorescence~\cite{KimbleMandel1976,NgChowKurtsiefer2022}. This antibunching measurement verifies the single-photon character of the fiber-coupled emission. Together, these measurements demonstrate 50 simultaneously operated, individually addressable and fiber-routed neutral-atom single-photon sources.

\vspace{6pt}
\noindent \textbf{Deterministic preparation and fast, sustained operation}

\noindent Large-scale atom arrays operated as single-photon sources face a central challenge in sustained operation. The stochastic loss of an atom prevents its corresponding channel from emitting further photons until the vacant trap is refilled, so uncompensated losses progressively reduce the number of active sources. We address this limitation by combining fast, channel-resolved atom discrimination with rapid feedback-controlled transport, which identifies vacancies and moves reserve atoms into the target sites~\cite{Pause2023,Pause2024,Gyger2024,Chiu2025}. This closed-loop recovery enables a large source array to operate for extended periods, as characterized in Fig.~\ref{Fig3}.

The waveguide interface gives each atom an independent, low-latency detection channel, allowing rapid discrimination of whether a trap is occupied~\cite{Chen2025Readout,MartinezDorantes2017,Tao2024,Hu2026}. Figure~\ref{Fig3}(a) shows the single-atom readout fidelity as a function of detection time. The 0-atom and 1-atom states are clearly separated in the photon-count distribution [inset of Fig.~\ref{Fig3}(a)]. For channel 6, a \SI{5}{\milli\second} detection window gives a readout fidelity of $98.6\%$, and the fidelity approaches unity for longer integration times. Because the background is dominated by detector dark counts, lower-noise detectors can further shorten the readout time at the same fidelity. Here the entire collected fluorescence is consumed by the occupancy measurement. In future operation, a second objective on the opposite side collects fluorescence into a disjoint solid angle, allowing occupancy verification and single-photon output to proceed through separate arms.

Since stochastic loading fills each trap with a probability of only $\sim0.6$~\cite{Schlosser2002,Schymik2022}, high-fill-fraction target arrays are prepared by real-time rearrangement~\cite{Barredo2016,Endres2016,Schymik2020,Tian2023,Pichard2024,Hu2026}. A first detection identifies the occupied traps, the atoms are transported to the target sites by reprogramming the AOD tones, and a second detection verifies the prepared array. Starting from an array of 48 traps [Fig.~\ref{Fig3}(b)], this procedure gives a per-site fill fraction of $96\%$ at $N=1$, decreasing to $93\%$ at $N=24$. The whole-array preparation success rate falls from $96\%$ at $N=1$ to $20\%$ at $N=24$, which sets the practical limit on the array size a single rearrangement cycle can deliver.

Reliable operation further requires that the prepared atoms remain available throughout the experimental sequence, which is addressed by repeating the detection during the hold and refilling emptied sites from a reservoir. Figure~\ref{Fig3}(c) compares the single-atom survival probability with and without such replenishment. Here rearrangement prepares a 10-atom target array within 24 traps, and the atoms occupying the remaining traps are retained as the reservoir. Without replenishment, the survival probability decays with a $1/e$ lifetime of $\approx\SI{1}{\second}$, limited by background-gas collisions and trap-induced heating; this measurement is performed with the probe light off. In the replenishment sequence, the periodic fluorescence detection required to identify emptied sites also laser-cools the atoms, and the combined effect of this cooling and the reservoir transfer extends the effective lifetime of an occupied source to $\approx\SI{10}{\second}$.

For a multiplexed photon source, the relevant figure of merit is not the lifetime of an individual atom but the fraction of time for which all channels are simultaneously occupied, since the loss of a single atom removes one channel from the array. This fraction is set by how quickly a loss is detected and repaired, which motivates characterizing the rate at which a defect-free array is restored. Figure~\ref{Fig3}(d) presents this array reset rate, defined as the rate at which a fresh target array is recovered by per-channel discrimination, rearrangement and reservoir replenishment. At small $N$ the reset rate reaches the detection-limited bound of \SI{118}{\hertz}, corresponding to a recovery time of $\sim\SI{7.5}{\milli\second}$ (detection and rearrangement only, excluding MOT preparation). The decrease with $N$ reflects the depletion of the reservoir. Each cycle consumes reservoir atoms to refill emptied target sites, and once the reservoir is exhausted the array can only be restored by reloading the MOT and repeating stochastic loading, which is far slower than a reservoir-based refill.

\vspace{6pt}
\noindent \textbf{Scaling the photonic interface toward 1,000 channels}

\noindent To examine the scalability of the interface, we fabricated a 784-channel chip using the same fan-out architecture. At the input facet, the waveguides form the compact 16-row by 49-column array shown in Fig.~\ref{Fig4}(a). At the output facet, the fan-out rearranges the channels into a $4\times196$ array with a \SI{127}{\micro m} pitch, allowing direct matching to a two-dimensional fiber array. We measured the transmittance of every channel and arranged the resulting insertion losses according to their physical positions in the 16-row by 49-column facet [Fig.~\ref{Fig4}(b)].

Excluding one channel with $T<1\%$, the per-channel insertion-loss distribution has a mean of \SI{6.4}{dB} and a channel-to-channel standard deviation of \SI{1.7}{dB} [Fig.~\ref{Fig4}(c)]. The excluded channel remains visible in the spatial map, where its loss saturates the color scale. The mean loss is higher than the \SI{2.9}{dB} measured for the 50-channel device, shown as a dashed reference in Fig.~\ref{Fig4}(c). The 784-channel chip measured here is a preliminary, unoptimized sample, and we expect its loss to be reduced to a comparable level as the fabrication is refined. Since the waveguide chip is laser-direct-written and terminates in a fiber-array-compatible geometry, the same interface concept can in principle be extended to source arrays of several thousands channels, provided that the trapped-atom array is enlarged accordingly.

\smallskip{}
\vspace{6pt}
\noindent \textbf{\large{}Discussion}{\large\par}

\noindent The central advance of this work is a scalable photonic interface that maps a \SI{5}{\micro\meter}-pitch neutral-atom tweezer array onto a standard \SI{127}{\micro\meter}-pitch fiber array. Across 50 individually addressed channels, the glass waveguide fan-out achieves an average insertion loss of \SI{2.9}{\decibel} and an average nearest-neighbor cross-talk of $0.4\%$. Unlike camera-based fluorescence imaging, it routes light from each atom into a dedicated waveguide, fiber and detector channel, thereby separating the microscopic geometry of the emitter array from the packaging requirements of downstream photonic systems. Combined with per-channel atom detection, rearrangement and reservoir replenishment, the interface enables source subarrays to be prepared with fill fractions above $93\%$ for target sizes up to 24 atoms, and restored at rates approaching the detection-limited bound of \SI{118}{\hertz}. These results establish the compatibility of channelized photon collection with the reconfigurability of optical-tweezer arrays.

A complementary use of the same photonic resource places the fiber array on the input side. Li \textit{et al.}\ established the same one-to-one channelization on the input side, generating each tweezer from its own single-mode fiber so that trapping and addressing light reach one atom through one channel~\cite{Li2025FiberArray}. Their approach couples the atoms directly to the fibers and thereby avoids the \SI{2.9}{dB} insertion loss of the waveguide array chip used here, while the present architecture offers higher potential scalability. In both cases the essential step is to decouple the channel geometry of the photonic interface from the geometry of the atom array. The two architectures differ mainly in where that decoupling is performed.

The main limitation is currently the photon collection efficiency, which is approximately $1\%$ into the waveguide mode and is primarily constrained by the collection objective and intermediate free-space optics. The measured antibunching, with $g^{(2)}(0)=0.29$ for one representative channel, verifies the single-photon character of the fiber-coupled emission, but does not establish uniform single-photon purity or mutual photon indistinguishability across the array. The most direct route to higher efficiency is a larger-numerical-aperture collection objective together with a shorter free-space path to the chip; beyond this, cavity-enhanced atom-photon interfaces~\cite{DuanKimble,Tiecke2014,Deist2022} or directional emission from collective Rydberg excitations~\cite{ParisMandoki2017,OrnelasHuerta2020,Zhang2025Repeater} could provide further gains, while lower-noise detectors would improve readout fidelity and reduce the atom-discrimination time. Establishing the platform as an array of mutually indistinguishable photon sources will further require array-wide measurements of brightness and $g^{(2)}(0)$, together with cross-channel spectral characterization or two-photon interference. Hong-Ou-Mandel interference~\cite{Hong1987} between channels was not attempted here, as the present collection efficiency would require prohibitively long integration times. Such measurements are deferred until the collection efficiency has been improved.

The fabrication of a waveguide chip containing 784 channels indicates that the fan-out architecture is not intrinsically limited to the 50-channel device demonstrated with atoms. Realizing a complete source system at this scale will additionally require a larger diffraction-limited field of view, increased trapping power, scalable detector and fiber packaging~\cite{Wollman2019}, and verification of loss and cross-talk uniformity across the full device. With these developments, the architecture could connect large, reconfigurable neutral-atom arrays directly to fiber networks and integrated photonic circuits~\cite{Ma2026Volcano,Zhang2025Repeater}. It therefore provides a scalable interface for parallel photon detection and correlation measurements today, and a route toward fiber-native single-photon source arrays for quantum networks and photonic quantum information processing.

\smallskip{}
\begin{acknowledgments}
We acknowledge Xu-Liang Zhang from Hangzhou Biaozhang Electronics for assistance. This work was funded by the National Key R\&D Program (Grant Nos. 2021YFA1402004 and 2021YFA1402002), the National Natural Science Foundation of China (Grant Nos.~92465201, T2325022 and U23A2074), and the Quantum Science and Technology-National Science and Technology Major Project (Grant Nos. 2021ZD0303200 and 2021ZD0301500). This work was also supported by  the Natural Science Foundation of Anhui
Province (Grant No.~2408085QA017), the Fundamental Research Funds for the Central Universities, USTC Research Funds of the Double First-Class Initiative. This work was partially carried out at the Supercomputing Center of USTC and the USTC Center for Micro and Nanoscale Research and Fabrication.
\end{acknowledgments}

\bibliographystyle{Zou}
\bibliography{references}

@article{Hu2026,
archivePrefix = {arXiv},
arxivId = {2607.08687},
author = {Hu, Ya-Dong and Ma, Dong-Qi and Zhang, Tian-Yang and Chen, Liang and Zhang, Yi-Chen and Zhong, Xiao-Kang and Zhu, Wen-Yi and Fan, Hong-Jie and Jie, Qing-Xuan and Zhang, Yan-Lei and Li, Gang and Ren, Xi-Feng and Zhang, Xu-Liang and Guo, Guang-Can and Wang, Zhu-Bo and Zou, Chang-Ling},
eprint = {2607.08687},
journal = {arXiv: 2607.08687},
month = {jul},
title = {{Low-latency FPGA-based electronic control system for fast preparation of defect-free atom arrays}},
url = {http://arxiv.org/abs/2607.08687},
year = {2026}
}

@Article{OBrien2009,
  author  = {O'Brien, Jeremy L. and Furusawa, Akira and Vu\v{c}kovi\'{c}, Jelena},
  title   = {Photonic quantum technologies},
  journal = {Nat. Photonics},
  volume  = {3},
  number  = {12},
  pages   = {687--695},
  year    = {2009},
  doi     = {10.1038/nphoton.2009.229}
}

@Article{Takeda2019,
  author  = {Takeda, S. and Furusawa, A.},
  title   = {Toward large-scale fault-tolerant universal photonic quantum computing},
  journal = {APL Photonics},
  volume  = {4},
  number  = {6},
  pages   = {060902},
  year    = {2019},
  doi     = {10.1063/1.5100160}
}

@Article{quantumsimulator,
  author  = {Aspuru-Guzik, Alán and Walther, Philip},
  title   = {Photonic quantum simulators},
  journal = {Nat. Phys.},
  volume  = {8},
  number  = {4},
  pages   = {285--291},
  year    = {2012},
  doi     = {10.1038/nphys2253}
}

@Article{Lindner2009,
  author  = {Lindner, Netanel H. and Rudolph, Terry},
  title   = {Proposal for Pulsed On-Demand Sources of Photonic Cluster State Strings},
  journal = {Phys. Rev. Lett.},
  volume  = {103},
  number  = {11},
  pages   = {113602},
  year    = {2009},
  doi     = {10.1103/PhysRevLett.103.113602}
}

@Article{Kimble2008,
  author  = {Kimble, H. J.},
  title   = {The quantum internet},
  journal = {Nature},
  volume  = {453},
  number  = {7198},
  pages   = {1023--1030},
  year    = {2008},
  doi     = {10.1038/nature07127}
}

@article{Giovannetti2011,
  author  = {Giovannetti, V. and Lloyd, S. and Maccone, L.},
  title   = {Advances in quantum metrology},
  journal = {Nat. Photonics},
  volume  = {5},
  pages   = {222--229},
  year    = {2011},
  doi     = {10.1038/nphoton.2011.35}
}

@Article{photonsourceReview,
  author  = {Caspani, Lucia and Xiong, Chunle and Eggleton, Benjamin J. and Bajoni, Daniele and Liscidini, Marco and Galli, Matteo and Morandotti, Roberto and Moss, David J.},
  title   = {Integrated sources of photon quantum states based on nonlinear optics},
  journal = {Light Sci. Appl.},
  volume  = {6},
  number  = {11},
  pages   = {e17100--e17100},
  year    = {2017},
  doi     = {10.1038/lsa.2017.100}
}

@Article{Lodahl2017,
  author  = {Lodahl, Peter and Mahmoodian, Sahand and Stobbe, S{\o}ren and Rauschenbeutel, Arno and Schneeweiss, Philipp and Volz, J{\"{u}}rgen and Pichler, Hannes and Zoller, Peter},
  title   = {Chiral quantum optics},
  journal = {Nature},
  volume  = {541},
  number  = {7638},
  pages   = {473--480},
  year    = {2017},
  doi     = {10.1038/nature21037}
}

@article{MeyerScott2020,
  author  = {Meyer-Scott, E. and Silberhorn, C. and Migdall, A.},
  title   = {Single-photon sources: Approaching the ideal through multiplexing},
  journal = {Rev. Sci. Instrum.},
  volume  = {91},
  pages   = {041101},
  year    = {2020},
  doi     = {10.1063/5.0003320}
}

@article{Kwiat1995,
  author  = {Kwiat, Paul G. and Mattle, Klaus and Weinfurter, Harald and Zeilinger, Anton and Sergienko, Alexander V. and Shih, Yanhua},
  title   = {New High-Intensity Source of Polarization-Entangled Photon Pairs},
  journal = {Phys. Rev. Lett.},
  volume  = {75},
  number  = {24},
  pages   = {4337--4341},
  year    = {1995},
  doi     = {10.1103/PhysRevLett.75.4337}
}

@article{Senellart2017,
  author  = {Senellart, P. and Solomon, G. and White, A.},
  title   = {High-performance semiconductor quantum-dot single-photon sources},
  journal = {Nat. Nanotechnol.},
  volume  = {12},
  pages   = {1026--1039},
  year    = {2017},
  doi     = {10.1038/nnano.2017.218}
}

@article{Aharonovich2016,
  author  = {Aharonovich, I. and Englund, D. and Toth, M.},
  title   = {Solid-state single-photon emitters},
  journal = {Nat. Photonics},
  volume  = {10},
  pages   = {631--641},
  year    = {2016},
  doi     = {10.1038/nphoton.2016.186}
}

@article{Beugnon2006,
  author  = {Beugnon, J. and Jones, M. P. A. and Dingjan, J. and Darqui{\'e}, B. and Messin, G. and Browaeys, A. and Grangier, P.},
  title   = {Quantum interference between two single photons emitted by independently trapped atoms},
  journal = {Nature},
  volume  = {440},
  pages   = {779--782},
  year    = {2006},
  doi     = {10.1038/nature04628}
}

@article{Schlosser2001,
  author  = {Schlosser, N. and Reymond, G. and Protsenko, I. and Grangier, P.},
  title   = {Sub-{P}oissonian loading of single atoms in a microscopic dipole trap},
  journal = {Nature},
  volume  = {411},
  pages   = {1024--1027},
  year    = {2001},
  doi     = {10.1038/35082512}
}

@article{Barredo2016,
  author  = {Barredo, D. and de L{\'e}s{\'e}leuc, S. and Lienhard, V. and Lahaye, T. and Browaeys, A.},
  title   = {An atom-by-atom assembler of defect-free arbitrary two-dimensional atomic arrays},
  journal = {Science},
  volume  = {354},
  pages   = {1021--1023},
  year    = {2016},
  doi     = {10.1126/science.aah3778}
}

@article{Endres2016,
  author  = {Endres, M. and Bernien, H. and Keesling, A. and Levine, H. and Anschuetz, E. R. and Krajenbrink, A. and Senko, C. and Vuleti{\'c}, V. and Greiner, M. and Lukin, M. D.},
  title   = {Atom-by-atom assembly of defect-free one-dimensional cold atom arrays},
  journal = {Science},
  volume  = {354},
  pages   = {1024--1027},
  year    = {2016},
  doi     = {10.1126/science.aah3752}
}

@article{Browaeys2020,
  author  = {Browaeys, A. and Lahaye, T.},
  title   = {Many-body physics with individually controlled {R}ydberg atoms},
  journal = {Nat. Phys.},
  volume  = {16},
  pages   = {132--142},
  year    = {2020},
  doi     = {10.1038/s41567-019-0733-z}
}

@article{Schymik2020,
  author  = {Schymik, K.-N. and Lienhard, V. and Barredo, D. and Scholl, P. and Williams, H. and Browaeys, A. and Lahaye, T.},
  title   = {Enhanced atom-by-atom assembly of arbitrary tweezer arrays},
  journal = {Phys. Rev. A},
  volume  = {102},
  pages   = {063107},
  year    = {2020},
  doi     = {10.1103/PhysRevA.102.063107}
}

@article{Pause2024,
  author  = {Pause, L. and Sturm, L. and Mittenb{\"u}hler, M. and Amann, S. and Preuschoff, T. and Sch{\"a}ffner, D. and Schlosser, M. and Birkl, G.},
  title   = {Supercharged two-dimensional tweezer array with more than 1000 atomic qubits},
  journal = {Optica},
  volume  = {11},
  pages   = {222--226},
  year    = {2024},
  doi     = {10.1364/OPTICA.513551}
}

@article{Darquie2005,
  author  = {Darqui{\'e}, B. and Jones, M. P. A. and Dingjan, J. and Beugnon, J. and Bergamini, S. and Sortais, Y. and Messin, G. and Browaeys, A. and Grangier, P.},
  title   = {Controlled single-photon emission from a single trapped two-level atom},
  journal = {Science},
  volume  = {309},
  pages   = {454--456},
  year    = {2005},
  doi     = {10.1126/science.1113394}
}

@article{Bluvstein2024,
  author  = {Bluvstein, D. and Evered, S. J. and Geim, A. A. and Li, S. H. and Zhou, H. and Manovitz, T. and Ebadi, S. and Cain, M. and Kalinowski, M. and Hangleiter, D. and Bonilla Ataides, J. P. and Maskara, N. and Cong, I. and Gao, X. and Sales Rodriguez, P. and Karolyshyn, T. and Semeghini, G. and Gullans, M. J. and Greiner, M. and Vuleti{\'c}, V. and Lukin, M. D.},
  title   = {Logical quantum processor based on reconfigurable atom arrays},
  journal = {Nature},
  volume  = {626},
  pages   = {58--65},
  year    = {2024},
  doi     = {10.1038/s41586-023-06927-3}
}

@Article{Tiecke2014,
  author  = {Tiecke, T. G. and Thompson, J. D. and de Leon, N. P. and Liu, L. R. and Vuleti\'{c}, V. and Lukin, M. D.},
  title   = {Nanophotonic quantum phase switch with a single atom},
  journal = {Nature},
  volume  = {508},
  pages   = {241--244},
  year    = {2014},
  doi     = {10.1038/nature13188}
}

@article{Dordevic2021,
  author  = {Dordevic, T. and Samutpraphoot, P. and Ocola, P. L. and Bernien, H. and Grinkemeyer, B. and Dimitrova, I. and Vuleti{\'c}, V. and Lukin, M. D.},
  title   = {Entanglement transport and a nanophotonic interface for atoms in optical tweezers},
  journal = {Science},
  volume  = {373},
  pages   = {1511--1514},
  year    = {2021},
  doi     = {10.1126/science.abi9917}
}

@article{Menon2024,
  author  = {Menon, S. G. and Glachman, N. and Pompili, M. and Dibos, A. and Bernien, H.},
  title   = {An integrated atom array--nanophotonic chip platform with background-free imaging},
  journal = {Nat. Commun.},
  volume  = {15},
  pages   = {6156},
  year    = {2024},
  doi     = {10.1038/s41467-024-50355-4}
}

@article{Li2025FiberArray,
  author  = {Li, X. and Hou, J.-Y. and Wang, J.-C. and Wang, G.-W. and He, X.-D. and Zhou, F. and Wang, Y.-B. and Liu, M. and Wang, J. and Xu, P. and Zhan, M.-S.},
  title   = {A fiber array architecture for atom quantum computing},
  journal = {Nat. Commun.},
  volume  = {16},
  pages   = {9728},
  year    = {2025},
  doi     = {10.1038/s41467-025-64738-8}
}

@article{Ma2026Volcano,
  author  = {Ma, D.-Q. and Jie, Q.-X. and Hu, Y.-D. and Zhu, W.-Y. and Zhang, Y.-C. and Fan, H.-J. and Zhong, X.-K. and Chen, G.-J. and Zhang, Y.-L. and Zhang, T.-Y. and Ren, X.-F. and Chen, L. and Wang, Z.-B. and Guo, G.-C. and Zou, C.-L.},
  title   = {Volcano architecture for scalable quantum processor units},
  journal = {Sci. China Phys. Mech. Astron.},
  volume  = {69},
  pages   = {220314},
  year    = {2026},
  doi     = {10.1007/s11433-025-2804-2}
}

@article{McKeever2004,
  author  = {McKeever, J. and Boca, A. and Boozer, A. D. and Miller, R. and Buck, J. R. and Kuzmich, A. and Kimble, H. J.},
  title   = {Deterministic generation of single photons from one atom trapped in a cavity},
  journal = {Science},
  volume  = {303},
  pages   = {1992--1994},
  year    = {2004},
  doi     = {10.1126/science.1095232}
}

@article{Davis1996,
  author  = {Davis, K. M. and Miura, K. and Sugimoto, N. and Hirao, K.},
  title   = {Writing waveguides in glass with a femtosecond laser},
  journal = {Opt. Lett.},
  volume  = {21},
  pages   = {1729--1731},
  year    = {1996},
  doi     = {10.1364/OL.21.001729}
}

@article{Marshall2009,
  author  = {Marshall, G. D. and Politi, A. and Matthews, J. C. F. and Dekker, P. and Ams, M. and Withford, M. J. and O'Brien, J. L.},
  title   = {Laser written waveguide photonic quantum circuits},
  journal = {Opt. Express},
  volume  = {17},
  pages   = {12546--12554},
  year    = {2009},
  doi     = {10.1364/OE.17.012546}
}

@article{Meany2015,
  author  = {Meany, T. and Gr{\"a}fe, M. and Heilmann, R. and Perez-Leija, A. and Gross, S. and Steel, M. J. and Withford, M. J. and Szameit, A.},
  title   = {Laser written circuits for quantum photonics},
  journal = {Laser Photonics Rev.},
  volume  = {9},
  pages   = {363--384},
  year    = {2015},
  doi     = {10.1002/lpor.201500061}
}

@article{Tan2021,
  author  = {Tan, D. and Wang, Z. and Xu, B. and Qiu, J.},
  title   = {Photonic circuits written by femtosecond laser in glass: improved fabrication and recent progress in photonic devices},
  journal = {Adv. Photonics},
  volume  = {3},
  pages   = {024002},
  year    = {2021},
  doi     = {10.1117/1.AP.3.2.024002}
}

@article{Corrielli2021,
  author  = {Corrielli, G. and Crespi, A. and Osellame, R.},
  title   = {Femtosecond laser micromachining for integrated quantum photonics},
  journal = {Nanophotonics},
  volume  = {10},
  pages   = {3789--3812},
  year    = {2021},
  doi     = {10.1515/nanoph-2021-0419}
}

@article{Hong1987,
  author  = {Hong, C. K. and Ou, Z. Y. and Mandel, L.},
  title   = {Measurement of subpicosecond time intervals between two photons by interference},
  journal = {Phys. Rev. Lett.},
  volume  = {59},
  pages   = {2044--2046},
  year    = {1987},
  doi     = {10.1103/PhysRevLett.59.2044}
}

@article{Kimble1977,
  author  = {Kimble, H. J. and Dagenais, M. and Mandel, L.},
  title   = {Photon antibunching in resonance fluorescence},
  journal = {Phys. Rev. Lett.},
  volume  = {39},
  pages   = {691--695},
  year    = {1977},
  doi     = {10.1103/PhysRevLett.39.691}
}

@article{Pause2023,
  author  = {Pause, L. and Preuschoff, T. and Sch{\"a}ffner, D. and Schlosser, M. and Birkl, G.},
  title   = {Reservoir-based deterministic loading of single-atom tweezer arrays},
  journal = {Phys. Rev. Res.},
  volume  = {5},
  pages   = {L032009},
  year    = {2023},
  doi     = {10.1103/PhysRevResearch.5.L032009}
}

@article{Gyger2024,
  author  = {Gyger, F. and Ammenwerth, M. and Tao, R. and Timme, H. and Snigirev, S. and Bloch, I. and Zeiher, J.},
  title   = {Continuous operation of large-scale atom arrays in optical lattices},
  journal = {Phys. Rev. Res.},
  volume  = {6},
  pages   = {033104},
  year    = {2024},
  doi     = {10.1103/PhysRevResearch.6.033104}
}

@article{Chiu2025,
  author  = {Chiu, N.-C. and Trapp, E. C. and Guo, J. and Abobeih, M. H. and Stewart, L. M. and Hollerith, S. and Stroganov, P. L. and Kalinowski, M. and Geim, A. A. and Evered, S. J. and Li, S. H. and Lyu, X. and Peters, L. M. and Bluvstein, D. and Wang, T. T. and Greiner, M. and Vuleti{\'c}, V. and Lukin, M. D.},
  title   = {Continuous operation of a coherent 3,000-qubit system},
  journal = {Nature},
  volume  = {646},
  pages   = {1075--1080},
  year    = {2025},
  doi     = {10.1038/s41586-025-09596-6}
}

@article{Chen2025Readout,
  author  = {Chen, L. and Zhu, W.-Y. and Chen, Z.-J. and Wang, Z.-B. and Hu, Y.-D. and Jie, Q.-X. and Guo, G.-C. and Zou, C.-L.},
  title   = {Optimized readout strategies for neutral-atom quantum processors},
  journal = {Phys. Rev. A},
  volume  = {112},
  pages   = {022606},
  year    = {2025},
  doi     = {10.1103/k2w2-83kc}
}

@article{MartinezDorantes2017,
  author  = {Martinez-Dorantes, M. and Alt, W. and Gallego, J. and Ghosh, S. and Ratschbacher, L. and V{\"o}lzke, Y. and Meschede, D.},
  title   = {Fast nondestructive parallel readout of neutral atom registers in optical potentials},
  journal = {Phys. Rev. Lett.},
  volume  = {119},
  pages   = {180503},
  year    = {2017},
  doi     = {10.1103/PhysRevLett.119.180503}
}

@article{Tao2024,
  author  = {Tao, R. and Ammenwerth, M. and Gyger, F. and Bloch, I. and Zeiher, J.},
  title   = {High-fidelity detection of large-scale atom arrays in an optical lattice},
  journal = {Phys. Rev. Lett.},
  volume  = {133},
  pages   = {013401},
  year    = {2024},
  doi     = {10.1103/PhysRevLett.133.013401}
}

@article{Schlosser2002,
  author  = {Schlosser, N. and Reymond, G. and Grangier, P.},
  title   = {Collisional blockade in microscopic optical dipole traps},
  journal = {Phys. Rev. Lett.},
  volume  = {89},
  pages   = {023005},
  year    = {2002},
  doi     = {10.1103/PhysRevLett.89.023005}
}

@article{Schymik2022,
  author  = {Schymik, K.-N. and Ximenez, B. and Bloch, E. and Dreon, D. and Signoles, A. and Nogrette, F. and Barredo, D. and Browaeys, A. and Lahaye, T.},
  title   = {In situ equalization of single-atom loading in large-scale optical tweezer arrays},
  journal = {Phys. Rev. A},
  volume  = {106},
  pages   = {022611},
  year    = {2022},
  doi     = {10.1103/PhysRevA.106.022611}
}

@article{Tian2023,
  author  = {Tian, W. and Wee, W. J. and Qu, A. and Lim, B. J. M. and Datla, P. R. and Koh, V. P. W. and Loh, H.},
  title   = {Parallel assembly of arbitrary defect-free atom arrays with a multitweezer algorithm},
  journal = {Phys. Rev. Appl.},
  volume  = {19},
  pages   = {034048},
  year    = {2023},
  doi     = {10.1103/PhysRevApplied.19.034048}
}

@article{Pichard2024,
  author  = {Pichard, G. and Lim, D. and Bloch, {\'E}. and Vaneecloo, J. and Bourachot, L. and Both, G.-J. and M{\'e}riaux, G. and Dutartre, S. and Hostein, R. and others},
  title   = {Rearrangement of individual atoms in a 2000-site optical-tweezer array at cryogenic temperatures},
  journal = {Phys. Rev. Appl.},
  volume  = {22},
  pages   = {024073},
  year    = {2024},
  doi     = {10.1103/PhysRevApplied.22.024073}
}

@Article{DuanKimble,
  author  = {Duan, L.-M. and Kimble, H. J.},
  title   = {Scalable Photonic Quantum Computation through Cavity-Assisted Interactions},
  journal = {Phys. Rev. Lett.},
  volume  = {92},
  pages   = {127902},
  year    = {2004},
  doi     = {10.1103/PhysRevLett.92.127902}
}

@article{Deist2022,
  author  = {Deist, E. and Lu, Y.-H. and Ho, J. and Pasha, M. K. and Zeiher, J. and Yan, Z. and Stamper-Kurn, D. M.},
  title   = {Mid-circuit cavity measurement in a neutral atom array},
  journal = {Phys. Rev. Lett.},
  volume  = {129},
  pages   = {203602},
  year    = {2022},
  doi     = {10.1103/PhysRevLett.129.203602}
}

@article{ParisMandoki2017,
  author  = {Paris-Mandoki, A. and Braun, C. and Kumlin, J. and Tresp, C. and Mirgorodskiy, I. and Christaller, F. and B{\"u}chler, H. P. and Hofferberth, S.},
  title   = {Free-space quantum electrodynamics with a single {R}ydberg superatom},
  journal = {Phys. Rev. X},
  volume  = {7},
  pages   = {041010},
  year    = {2017},
  doi     = {10.1103/PhysRevX.7.041010}
}

@article{OrnelasHuerta2020,
  author  = {Ornelas-Huerta, D. P. and Craddock, A. N. and Goldschmidt, E. A. and Hachtel, A. J. and Wang, Y. and Bienias, P. and Gorshkov, A. V. and Rolston, S. L. and Porto, J. V.},
  title   = {On-demand indistinguishable single photons from an efficient and pure source based on a {R}ydberg ensemble},
  journal = {Optica},
  volume  = {7},
  pages   = {813--819},
  year    = {2020},
  doi     = {10.1364/OPTICA.391485}
}

@article{Zhang2025Repeater,
  author  = {Zhang, Y.-L. and Jie, Q.-X. and Li, M. and Wu, S.-H. and Wang, Z.-B. and Zou, X.-B. and Zhang, P.-F. and Li, G. and Zhang, T. and Guo, G.-C. and Zou, C.-L.},
  title   = {Architecture for a quantum repeater based on {R}ydberg-atom quantum processors},
  journal = {Phys. Rev. Appl.},
  volume  = {24},
  pages   = {024052},
  year    = {2025},
  doi     = {10.1103/8rss-rqr2}
}

@article{Wollman2019,
  author  = {Wollman, E. E. and Verma, V. B. and Lita, A. E. and Farr, W. H. and Shaw, M. D. and Mirin, R. P. and Nam, S. W.},
  title   = {Kilopixel array of superconducting nanowire single-photon detectors},
  journal = {Opt. Express},
  volume  = {27},
  pages   = {35279--35289},
  year    = {2019},
  doi     = {10.1364/OE.27.035279}
}

@article{KimbleMandel1976,
  author  = {Kimble, H. J. and Mandel, L.},
  title   = {Theory of Resonance Fluorescence},
  journal = {Physical Review A},
  year    = {1976},
  volume  = {13},
  number  = {6},
  pages   = {2123--2144},
  month   = jun,
  doi     = {10.1103/PhysRevA.13.2123}
}

@article{NgChowKurtsiefer2022,
  author  = {Ng, Boon Long and Chow, Chang Hoong and Kurtsiefer, Christian},
  title   = {Observation of the {Mollow} Triplet from an Optically Confined Single Atom},
  journal = {Physical Review A},
  year    = {2022},
  volume  = {106},
  number  = {6},
  pages   = {063719},
  month   = dec,
  doi     = {10.1103/PhysRevA.106.063719}
}

\end{document}